\documentclass[aps,prl,showpacs,floatfix,twocolumn,amsmath,amssymb,nofootinbib]{revtex4-1}

\usepackage{amssymb,amsfonts,amsmath}
\usepackage{subfig}
\usepackage[pdftex]{graphicx}

\usepackage{amssymb,amsfonts,amsmath}

\usepackage[dvipsnames]{xcolor}

\newcommand{\beq}{\begin{equation}}
\newcommand{\beqn}{\begin{eqnarray}}
\newcommand{\eeq}{\end{equation}}
\newcommand{\eeqn}{\end{eqnarray}}

\begin{document}




\title{Formation of Scientific Fields as a Universal Topological Transition}





\author{Lu\'is M. A. Bettencourt$^{1}$ and David I. Kaiser$^{2}$}

\affiliation{$^1$ Santa Fe Institute, Santa Fe, NM USA \\
$^2$ Massachusetts Institute of Technology, Cambridge, MA USA} 

\email{Email addresses: bettencourt@santafe.edu ,  dikaiser@mit.edu}

\date{\today}




\begin{abstract} 
Scientific fields differ in terms of their subject matter, research techniques, collaboration sizes, rates of growth, and so on. We investigate whether common dynamics might lurk beneath these differences, affecting how scientific fields form and evolve over time. Particularly important in any field's history is the moment at which shared concepts and techniques allow widespread exchange of ideas and collaboration. At that moment, co-authorship networks show the analog of a percolation phenomenon, developing a giant connected component containing most authors. We develop a general theoretical framework for analyzing finite, evolving networks in which each scientific field is an instantiation of the same large-scale topological critical phenomenon. We estimate critical exponents associated with the transition and find evidence for universality near criticality implying that, as various fields approach the topological transition, they do so with the same set of critical exponents consistent with an effective dimensionality $d \simeq 1$. These results indicate that a common dynamics is at play in all scientific fields, which in turn may hold policy implications for ways to encourage and accelerate the creation of scientific and technological knowledge. 
\vskip 8pt
SFI Working Paper 2015-03-009, Preprint MIT-CTP-4652
\end{abstract}


\preprint{MIT-CTP 4652}

\maketitle





\section{Introduction} 

The evolution of science and technology is a subject of enormous intellectual and societal importance~\cite{kuhn,price}. One inescapable feature in recent decades has been the explosion of scientific publishing worldwide, along with enormous growth in the indexing and availability of scholarly documents~\cite{Tabah99}. These developments open new opportunities to construct a robust, quantitative understanding of the processes by which scientific fields emerge and grow, and how the dynamics of knowledge creation and diffusion may be encouraged by policy. 

A central question is whether there exist common structures and dynamics in the evolution of science that govern the development of different fields across disciplines and time. Foundational work toward a general science of science~\cite{kuhn,price} assumed that different fields reflect, at least in some general sense, a common dynamics of discovery. Quantitative methods with which to explore such dynamics have included population dynamical models, 
networks of co-citation and collaboration,
disciplinary maps of science, 
and phylogenetic term analyses, 
among others. (For reviews, see \cite{Tabah99,AlbertBarabasi2002,Dorogovtsev2002,Castellano09,Newman10,Kuperman10,Boerner10}.)

These studies leave open the question as to whether common dynamics are shared across scientific fields. It is clear, for example, that fields of science differ in terms of time scales of growth, levels of investment, necessary equipment, collaboration size, and scientific productivity, as standard measures of scientific impact have begun to show~\cite{Redner05,H-index}. On one end, small and relatively clearly defined fields exist, such as the study of cosmic strings in theoretical physics, with a few thousand authors, which are characterized by shared concepts and techniques. On the other end, much larger fields coalesce, such as nanotechnology, with hundreds of thousands of authors, which show much more diversity in topic and method and whose collaboration networks are more loosely connected. In the face of such differences, do any commonly shared features remain?

In this paper we develop an integrated analysis of several fields of science spanning a wide range of disciplines, methods, and sizes, from theoretical physics to computer science and biomedical research. We show that there exists a strong set of commonalities across all these fields, signaling the advent of cohesive communities of practice. These changes in social dynamics and cognitive content are manifest in terms of a topological transformation of each field's large-scale structure of collaboration. In fact, all fields analyzed here can be interpreted as undergoing the same kind of idealized phase transition: a topological transition (akin to percolation) after which nearly all authors active in the field may be connected to all others by a finite number of co-authorship links.

To analyze these topological transitions, we develop a theoretical framework based on universality classes of critical exponents in the study of critical phenomena, familiar from statistical mechanics~\cite{Fisher1971,ZinnJustin,Binder}. In phase transitions in matter, such as a liquid boiling to a gas, common properties persist across different materials: though water, carbon dioxide, and liquid helium boil at different critical temperatures, they approach the liquid-gas phase transition in exactly the same way, governed by the same universal dynamics. In particular, they share the same ``critical exponents" near the transition: exponents that determine how quickly a given system approaches criticality. In physical systems, the existence of shared critical exponents across different substances indicates that the physical nature of the transition is independent of microphysical details specific to any given substance. 

We extend this framework to study topological transitions in co-authorship networks of scientific fields. In making this extension, we introduce two important modifications. First, unlike the usual treatment of physical systems, our scientific networks are always finite in size. We can never take the thermodynamic limit in which the number of nodes, $N$, grows to infinity, but rather must work in the regime $0 < N < \infty$. Second, the size of our networks changes over time, $dN / dt \neq 0$. Hence we introduce various scaling relationships that take into account the finite, evolving nature of these networks. Using those scaling relationships, we may isolate shared features across scientific fields that would otherwise remain masked by their differences in size and rates of growth. In this way, we may demonstrate that in a very specific quantitative sense, all fields are comparable under scale transformations. Each field that we study here maps to an idealized dynamical critical phenomenon that can exist at any scale, including the limit of very large numbers of authors. This idealized transition corresponds to a specific set of values for the critical exponents, consistent with an effective dimensionality for these networks near criticality, $d \simeq 1$.

\section{Results}
\subsection{Characterizing Growth Over Time}

In this study we have examined the growth and development of several scientific and technical fields as they changed over a time-scale of decades. The fields vary greatly in size and composition: from relatively modest-sized communities in theoretical physics such as cosmic strings or cosmological inflation, in which authors have similar training; to benchtop biomedical topics like research on scrapie and prions, which incorporate co-authors of varied expertise who work together on a narrowly-defined problem; to huge interdisciplinary fields like nanotechnology and sustainability science, which feature authors from a wide range of specialties. See Table 1.
\beq
\nonumber
\begin{array}{cccc} 
{\rm field} & {\rm years} & {\rm publications} & {\rm authors} \\
{\rm scrapie \> \& \> prions} & 1960-2005 & 11,074 & 14,620 \\
{\rm quantum \> computing} & 1967-2005 & 8,946 & 7,518 \\
{\rm string \> theory} & 1974-2005 & 9,766 & 25,022 \\
{\rm sustainability \> science} & 1974-2009 & 20,455 & 36,984 \\
{\rm cosmic \> strings} & 1976-2005 & 2,443 & 2,292 \\
{\rm inflation} & 1981-2005 & 5,135 & 3,410 \\
{\rm H5N1 \> influenza} &  1988-2011 & 86,885 & 66,629 \\
{\rm nanotechnology} & 1990-2010 & 521,075 & 333,990 \\
{\rm carbon \> nanotubes} &  1992-2011 & 13,500 & 4,190 \\
\end{array}
\label{table1}
\eeq
\centerline{\small {\bf Table 1}. Summary statistics for nine scientific fields.}
\vskip 10pt

In our previous work we have found two features of scientific authorship that can be used to simplify our analysis. First, as our mean-field population-modeling has shown, the total number of authors (nodes), $N (t)$, plays the role of the relevant time-like dynamical variable \cite{BKKCW08}. For example, we found that the total number of articles grows, on average, as a simple power-law of the total number of authors, even as both quantities display more complicated growth patterns when measured with respect to time, $t$. Thus we may adopt $N (t)$ as our time-like variable, akin to using scale factor, $a (t)$, rather than time, $t$, in cosmological studies of the dynamics of the Universe.

Second, in our previous study of collaboration networks \cite{BKK09}, we found a simple scaling relationship between the number of co-authorship links (or edges) per node, $E(t)$, and the number of authors, $N (t)$,
\begin{eqnarray}
E (t) = E_i \left[ N (t) \right]^{\alpha_i} ,
\label{ENalpha}
\end{eqnarray}
where the constants $E_i$ and $\alpha_i$ differ among the scientific fields, $i$, under consideration. The quantity $\alpha$ is known as the ``densification exponent": networks with $\alpha > 1$ grow more dense as they evolve over time \cite{Leskovec2005}.  Virtually every field we studied in \cite{BKK09} had $\alpha > 1$, with most falling in the range $\alpha = \left[ 1.05 , 1.38 \right]$; only cold fusion displayed a scaling exponent consistent with $1$. The largest fields in our sample have comparable densification exponents: $\alpha = 1.27$ for sustainability science, and $\alpha = 1.36$ for nanotechnology. (See Figure 1.)

Such densification over time provides one simple example of the clarity that can come from considering $N$ to be the time-like variable, rather than ordinary time, $t$. Consider, for example, the plots in Figure 1. Panels A and B show the number of new authors per year in nanotechnology and sustainability science, respectively. For nanotechnology, new authors grew roughly exponentially over much of the time period of interest, whereas new authors in sustainability science grew even faster. Yet when plotting the number of links in each network, $E(t)$, versus number of authors, $N(t)$, as in panels C and D, both fields betray the simple scaling behavior of Eq. (\ref{ENalpha}) --- a simplicity of structure that is not at all apparent from time-series graphs like those in panels A and B.

\begin{figure}
   \includegraphics[width=3.5in]{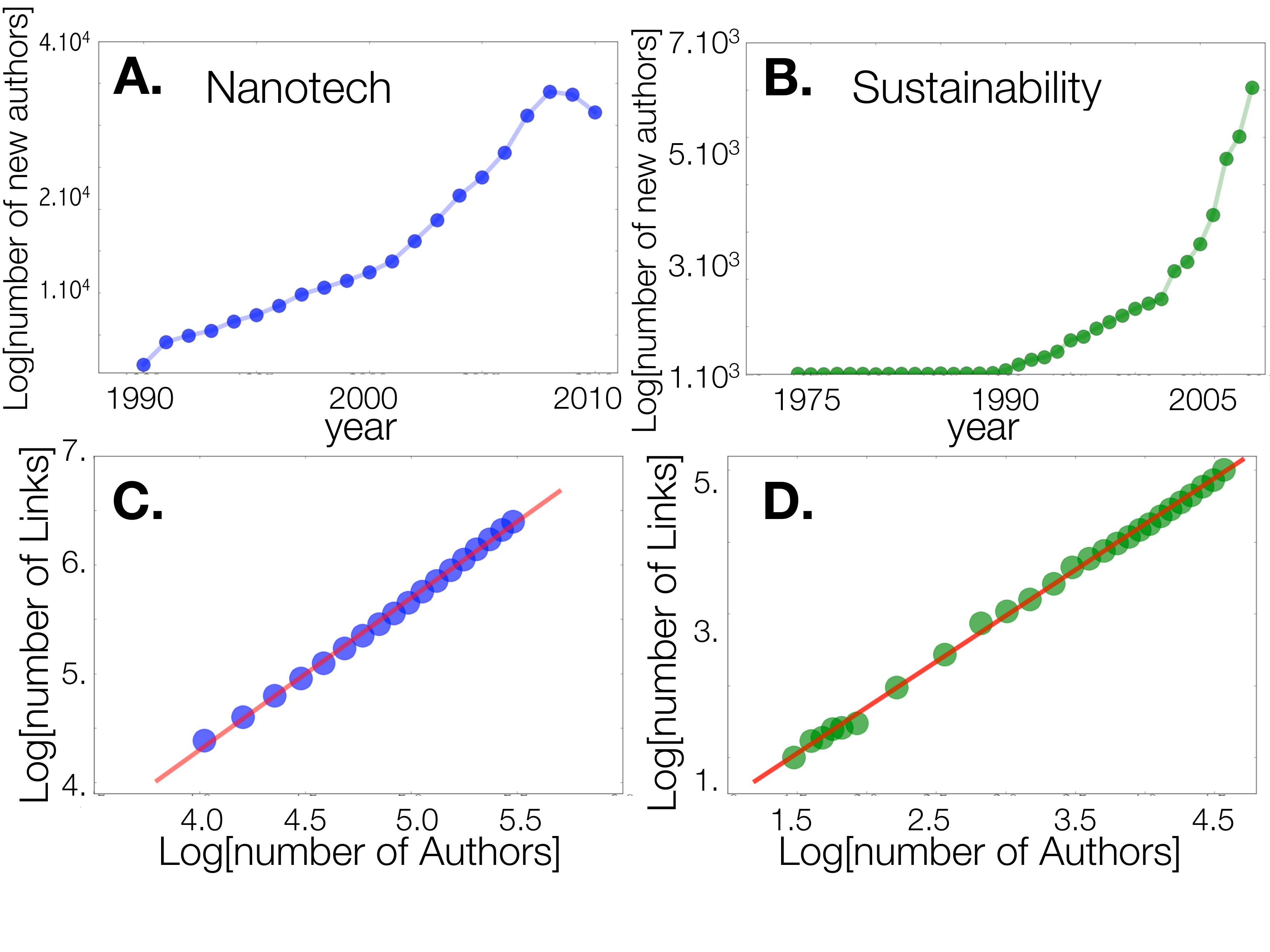} 
   \caption{\small Time evolution of scientific fields and the densification of collaboration networks. A. The number of new authors per year in nanotechnology vs. time. B. The number of new authors per year  in sustainability science vs. time. Regardless of the details of the temporal evolution of the field, the number of co-authorship links increases in a scale-invariant way with the number of authors in the network, as shown in C for nanotechnology and D for sustainability science. These patterns are similar to those we identified in \cite{BKK09} for a variety of smaller scientific fields.}
   \label{fig:1}
\end{figure}

\subsection{From Densification to Percolation}

A quantity of interest in any network study is the degree of a node, $k$, which counts the number of links connected to that node. Given the form of Eq. (\ref{ENalpha}), the average degree, $\langle k \rangle$, should obey a simple scaling relation with network size, $N$, of the form
\beq
\langle k \rangle =2 \frac{E (t) }{N(t)} = 2 E_i \left[ N (t) \right]^{\alpha_i - 1} .
\label{kavg}
\eeq
In \cite{BKK09} we found that fields with exponents $\alpha > 1$ underwent a topological transition, akin to percolation: at some finite time in each field's evolution, one giant connected cluster emerged such that nearly all nodes were connected to the cluster by a finite number of steps. In other words, following the topological transition, nearly every author in the scientific field could be connected to any other author by a finite set of co-authorship links: author $A$ wrote a paper with author $B$, who independently wrote a different paper with author $C$, and so forth, until the set of authors became fully connected. 

Such topological transitions signal an important step in a scientific field's development: a distinct, robust scientific field seems to emerge only once there exists some commonly shared set of research questions, concepts, and methods that allow multiple authors to cooperate and collaborate. The topological transition in co-authorship networks, in other words, might provide a signal --- available to policymakers and scientists alike --- that a new topic has emerged into a full-fledged field of inquiry. Such critical behavior occurs at different times in various fields' evolution, and hence at widely varying sizes of networks. If one relied only upon time-series data such as new authors or publications per year, one would miss the underlying similarities in topology and network structure. 

To explore the dynamics of these topological transitions, we analyze the co-authorship networks using techniques similar to those developed for the study of phase transitions in physical systems~\cite{ZinnJustin}. Phase transitions are system-wide structural  changes in the vicinity of a critical point, such as the boiling point for a liquid-to-gas transition, or the Curie temperature above which a ferromagnet loses its magnetization. In these examples, the critical parameter, $\tau$, depends on the ratio of the system's temperature to its critical value: $\tau = [ (T / T_c) - 1]$. In our previous work \cite{BKK09}, we found suggestive evidence that fields with $\alpha > 1$ might percolate when their average degree, $\langle k \rangle$, crosses some field-dependent critical threshold, $k_c$. That is, percolation might occur once the average connectivity of a network, changing in time as in Eq. (\ref{kavg}), reaches a critical point. Such behavior would be compatible with previous observations of percolation on random graphs \cite{AlbertBarabasi2002,Dorogovtsev2002,Castellano09,Newman10}. We therefore define the critical parameter as
\beq
\tau \equiv \left[ \frac{ \langle k \rangle}{k_c} - 1 \right] 
\label{tau1}
\eeq
and study the dynamics of a given field in the vicinity of $\vert \tau  \vert \sim 0$. Given the form of Eq. (\ref{kavg}), we may rewrite Eq. (\ref{tau1}) as
\beq
\tau = \left[ \left( \frac{N}{N_c} \right)^{\alpha_i - 1} - 1 \right] ,
\label{tau2}
\eeq
where $N_c$ is the size of the network at criticality. Using Eq. (\ref{kavg}), we find $k_c = 2 E_i \left[ N_c \right]^{\alpha_i - 1}$. Like a freezing point, the critical degree $k_c$ (and hence $N_c$) will vary by field, see Table~2. 
\beq
\nonumber
\begin{array}{cccccc}
{\rm field} & E_i & \alpha_i & N_c  & k_c^{\rm e} \> & k_c^{\rm m} \>  \\
{\rm scrapie \> \& \> prions} & 1.53 & 1.12 & 61 & 5.01 & 2.03 \\
{\rm quantum \> computing} & 0.38 & 1.22 & 171 & 2.36 & 2.85  \\
{\rm string \> theory} & 0.09 & 1.36 & 915 & 2.10 & 2.00  \\
{\rm sustainability \> science} & 0.15 & 1.27 & 6285 & 3.18 & 3.05 \\
{\rm cosmic \> strings} & 0.28 & 1.21 &  228 & 1.75 & 2.16 \\
{\rm inflation} & 0.09 & 1.38 & 114 & 1.09 & 1.65  \\
{\rm H5N1\>  influenza} & 3.22 & 1.05 & 18 & 7.44 & 3.67  \\
{\rm nanotechnology} & 0.07&  1.36 & 1460  & 1.93 & 3.72  \\
{\rm carbon \>  nanotubes} & 0.93 & 1.17 & 157 & 4.39 & 4.01  \\
\end{array}
\label{table2}
\eeq
\centerline{\small {\bf Table 2}. Links-per-node, densification exponents, critical} 
\centerline{\small network size, and critical degree (expected and measured)}
\centerline{\small  for the fields of Table 1.}
\vskip 10pt

In general, Eq. (\ref{ENalpha}), with which we define the densification exponent $\alpha_i$, provides an excellent fit to the data. For example, the fits in Figures 1C and 1D each have $R^2 = 0.998$. But the scaling relation of Eq. (\ref{ENalpha}) tends to underestimate the number of links per node for small networks (or, equivalently, at early times in the evolution of large networks), and hence tends to underestimate $\langle k \rangle$ near criticality, at $N = N_c$. The values of $k_c^{\rm e} = 2 E_i \left[ N_c \right]^{\alpha_i - 1}$ therefore often differ from direct measures of $k_c$ at criticality, $k_c^{\rm m}$. 

The behavior of the system in the vicinity of $\vert \tau \vert \sim 0$ may be characterized by several quantities that scale with $\tau$. The first is the percolation probability, $P (\tau)$: the likelihood that a randomly-selected node belongs to the largest connected cluster, which may be computed easily as the fraction of nodes in the largest connected cluster. $P(\tau) $ is an order parameter for the system, akin to bulk magnetization for a lattice of spins in a ferromagnet placed in an external magnetic field. The second quantity is the susceptibility per node, $ S (\tau) = \sum_s s^2 n_s/N$, where $n_s$ is the number of clusters that contain $s$ nodes. (The sum extends over all clusters except the largest connected cluster.) $S(\tau)$ thus characterizes the variance of fluctuations per node in cluster size for the system, which is highest at  the onset of the formation of a giant graph component, at the critical point. A final quantity of interest is the correlation length $\xi (\tau)$, which characterizes how smooth or homogenous the system is. Effectively, $\xi (\tau)$ measures the size of the largest clusters that are not part of the single largest component \cite{ZinnJustin}.
In the infinite-volume limit ($N \rightarrow \infty$), these quantities scale near $\vert \tau \vert \sim 0$ as
\beq
P (\tau) \propto \vert \tau \vert^\beta , \quad S (\tau) \propto \vert \tau \vert^{-\gamma} , \quad \xi (\tau) \propto \vert \tau \vert^{-\nu} ,
\label{PSxiInfinite}
\eeq
in terms of the so-called critical exponents, $\beta$, $\gamma$, and $\nu$, with $\beta, \> \gamma, \> \nu \geq 0$ \cite{ZinnJustin}. In the limit $N \rightarrow \infty$, as the system passes through the critical point, the percolation probability $P$ should rise from zero as it becomes more and more likely that any given node will belong to the largest connected cluster. The susceptibility $S$ and correlation length $\xi$ should each fall from large (divergent) values, as fewer nodes remain in clusters separate from the largest connected cluster. In a finite system, however, the correlation length can only grow as large as the system itself: $\xi \leq N$. Likewise, in a finite system clusters can only grow so large, so $S$ will reach some maximum but finite value near $\vert \tau \vert \sim 0$, rather than diverging to infinity; while $P$ will remain at some small but nonzero value near $\vert \tau \vert \sim 0$. See Figure 2.
\begin{figure}
   \centering
   \includegraphics[width=3.5in]{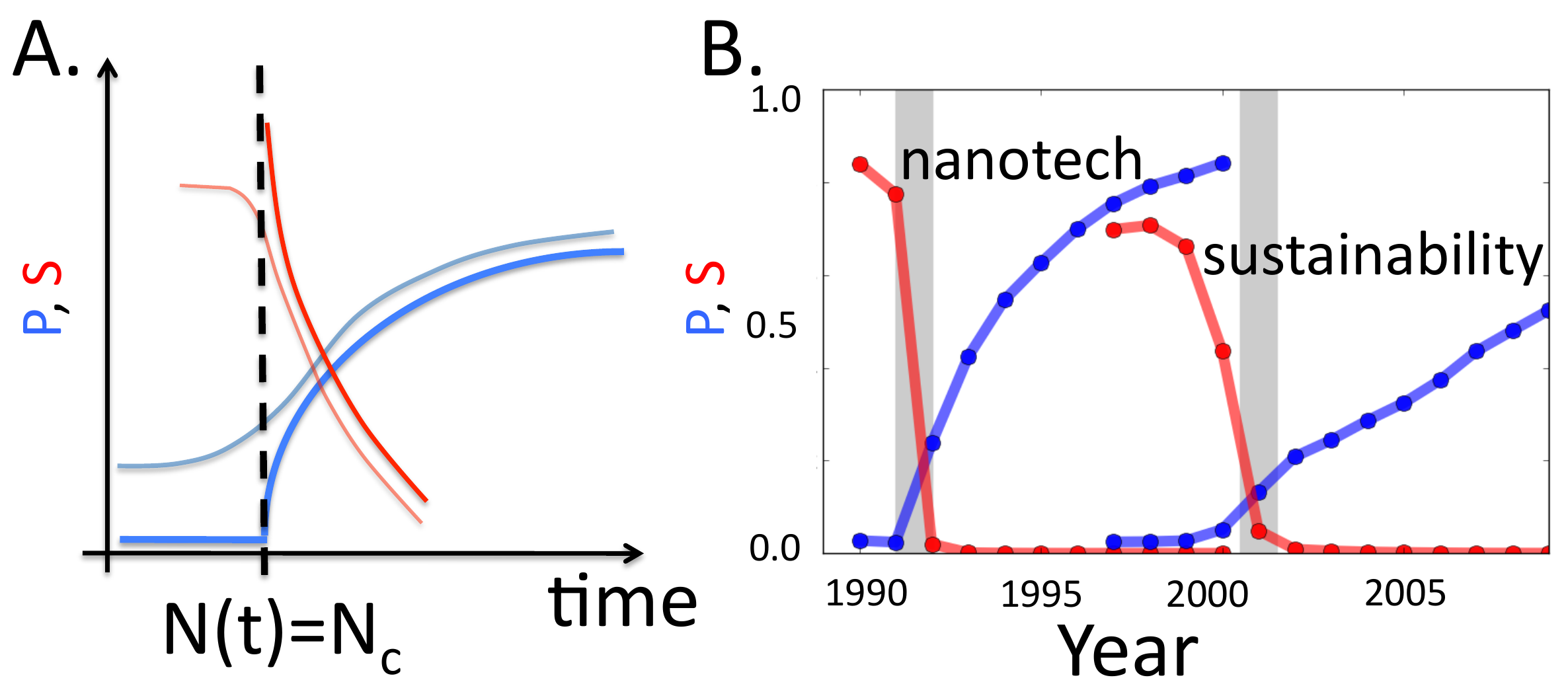} 
   \caption{\small Signatures of the topological transition for the formation of scientific fields. A. Schematic representation of the cumulants $P$ (blue) and $S$ (red). In the idealized infinite-size limit, a field undergoes a topological transition such that $P$ grows away from zero while $S$ quickly decays from a large value. In this limit, $P$ has a derivative discontinuity at the critical point, while $S$ diverges at the critical point. The behavior of these quantities for real fields are finite-size approximations (pale lines) to their infinite-size extrapolations (bright lines). B. The actual critical behavior for nanotechnology and sustainability science. }
   \label{fig:2}
\end{figure}

Remarkably, many physical systems display ``universality": though they boil at different critical temperatures, $T_c$, they all boil in the same way, that is, they share the same values for the set of critical exponents. Systems that share the same values of $\beta$, $\gamma$, and $\nu$ fall into the same {\it universality class}. The power of these scaling relationships thus becomes clear: systems that might have quite different microscopic properties and dynamics each behave in {\it exactly the same way} near a phase transition. The values of the critical exponents depend on the dimensionality of the system, $d$, and obey a ``hyperscaling" relation \cite{ZinnJustin,Binder,Henkel} 
\beq
d \cdot \nu = \gamma + 2 \beta.
\label{hyperscaling}
\eeq

In order to account for the finite size of our systems, we make use of the usual scaling hypothesis \cite{Fisher1971,ZinnJustin,Binder,Henkel}. At any given time, the only relevant length scales are the smallest scale in the system (an individual author), the total number of authors $N$, and the correlation length $\xi$. From Eq. (\ref{PSxiInfinite}) we see that near the critical point, $\vert \tau \vert \propto \xi^{-1/\nu}$. Since we assume that the only dimensionful quantity of relevance to the dynamics is $\tau$, we scale the quantities $P$ and $S$ as
\beq 
P (\tau) = \left[ \xi (\tau) \right]^{-\beta / \nu} P_0 (x) , \quad S (\tau) = \left[ \xi (\tau) \right]^{\gamma / \nu} S_0 (x) ,
\label{PSfinite1}
\eeq
where the quantities $P_0$ and $S_0$ depend only on the dimensionless ratio $x \equiv N / \xi$. We make a further rescaling to extract the dependence on the system size $N$ by writing $P_0 = x^{-\beta / \nu} \tilde{P} (x)$ and $S_0 = x^{\gamma / \nu} \tilde{S} (x)$. In order for $P$ and $S$ to scale with $\tau$ as in Eq. (\ref{PSxiInfinite}), we expect $\tilde{P}$ and $\tilde{S}$ to scale as functions of $x^{1/\nu} \propto N^{1/\nu} \vert \tau \vert$. We may then write Eq. (\ref{PSfinite1}) as
\beq
P (\tau) = P^* (N) \>  f (N^{1/\nu} \tau ) , \quad S (\tau) = S^* (N) \>  g (N^{1/\nu} \tau ) 
\label{PSfinite3}
\eeq
in terms of the functions
\beq
P^* (N) \equiv C_0 N^{- \beta / \nu} , \quad S^* (N) \equiv D_0 N^{\gamma / \nu} ,
\label{PstarSstar}
\eeq
where $C_0$ and $D_0$ are constants. We normalize $f (0) = g(0) = 1$ at the critical point, $\tau = 0$. 
Thus we find that near the critical point, as the correlation length rapidly grows to become comparable to the size of the entire system ($\xi \rightarrow N$), the percolation probability and susceptibility for a finite system scale as $P \rightarrow P^* (N_c)$ and $S \rightarrow S^* (N_c)$. 

More generally, as we see from Eq. (\ref{PSfinite3}), away from the critical point we may plot $P / P^*$ versus $N^{1/\nu} \tau$ to yield a single curve, $f$, on which all members of the same universality class should fall. Likewise, plotting $S / S^*$ versus $N^{1/\nu} \tau$ should yield a single curve, $g$, on which all members fall.  The form of these curves is generally expected to be scale invariant (power-law) away from the critical point, but must also obey the boundary conditions $f(0)=g(0)=1$.  We may satisfy both these conditions with the ansatz
\begin{eqnarray}
f(y) = (1+ a_f y)^\beta , \qquad g(y) = \frac{1}{(1+a_g y)^{\gamma}}  ,
\label{fgdef}
\end{eqnarray} 
where $y  \equiv N^{1/\nu} \vert \tau \vert \propto x^{1/\nu}$, and $a_f$ and $a_g$ are constants.

Our empirical results are consistent with the hypothesis of universality. First consider Figure 3, which shows the scaling of $P^*$ and $S^*$ for each scientific field at criticality, $N = N_c$. All nine fields cluster around a single line for $P^*$ and $S^*$, with exponents given by $\beta / \nu = 0.46$ (95\% confidence interval $[0.22,0.69]$, $R^2=0.69$) and $\gamma / \nu = 0.19$ (95\% confidence interval $[0.12,0.25]$, $R^2=0.82$). The constant coefficients take the values $C_0 = 1.21$ and $D_0 = 1.69$.

\begin{figure}
   \centering
   \includegraphics[width=3.5in]{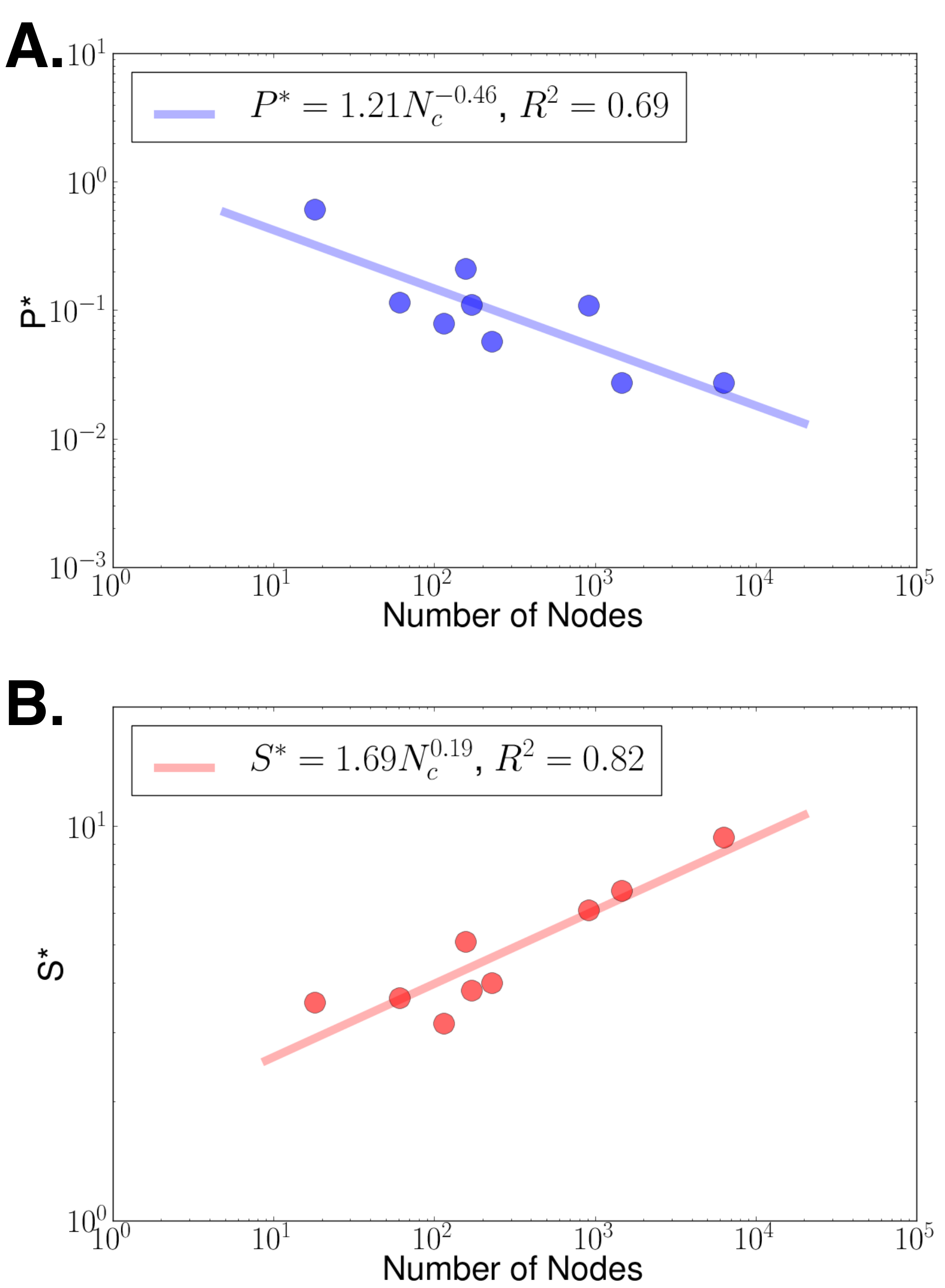}
   \caption{\small The finite-size scaling of $P^*$ and $S^*$ at criticality for nine fields with different sizes, with $P^*$ and $S^*$ given in Eq. (\ref{PstarSstar}). A. The percolation probability, $P^*$, decays to zero as a scale-invariant function of $N_c$ with exponent $\beta / \nu = 0.46$ (95\% confidence interval $[0.22,0.69]$). B. The susceptibility, $S^*$, diverges to infinity with increasing $N_c$, with an exponent $\gamma / \nu = 0.19$ (95\% confidence interval $[0.12,0.25]$). The behavior of both these quantities suggests the existence of an infinite-size transition of which all fields in our sample are finite-sized realizations. }
   \label{fig:3}
\end{figure}

\begin{figure}
   \centering
   \includegraphics[width=3.5in]{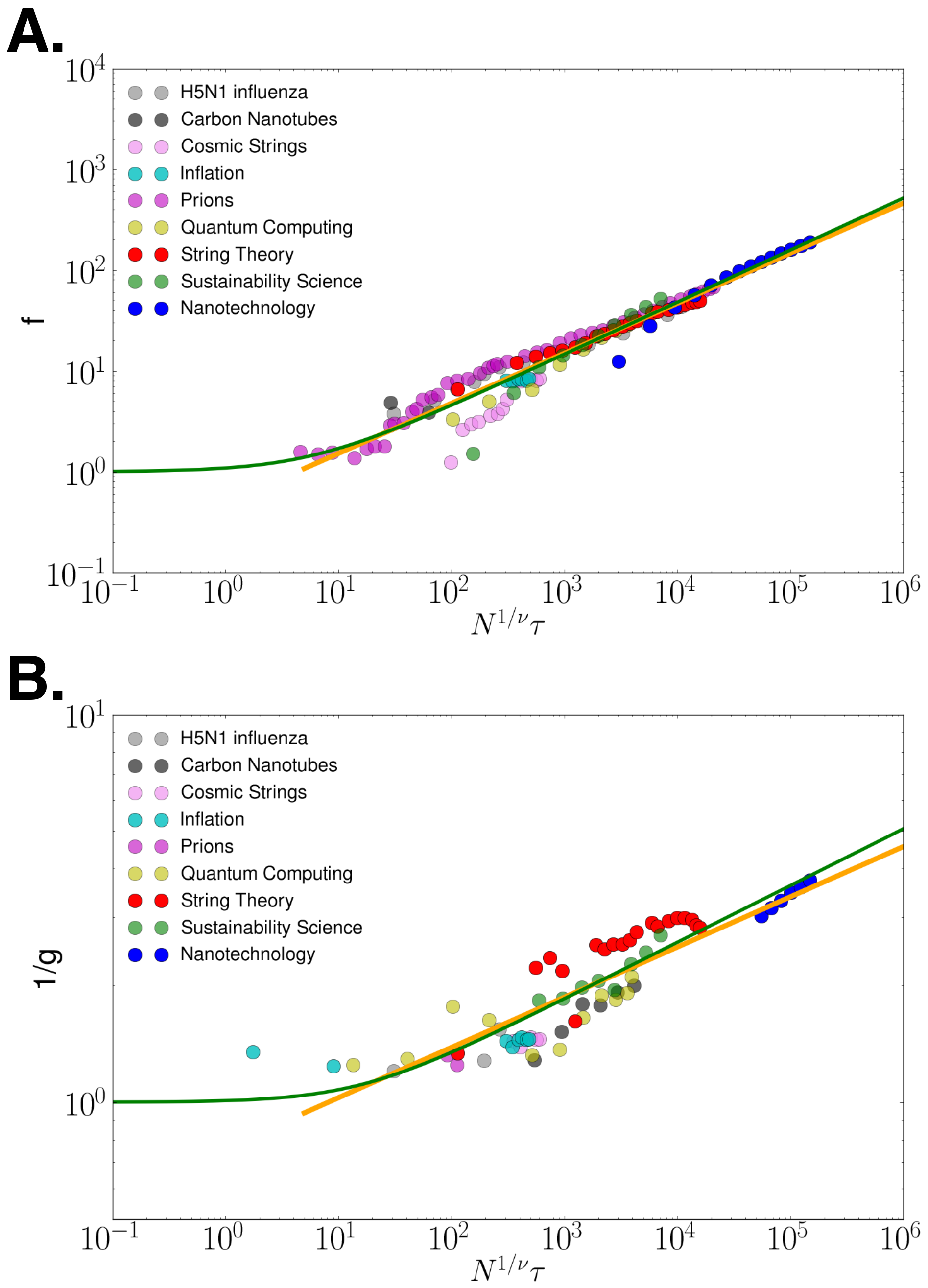} 
   \caption{\small The behavior of the percolation cumulants $P$ and $S$ for the nine scientific fields. In both plots, the yellow lines show the expected power-law scalings for best-fit values of the critical exponents. The green line shows a nonlinear fit to the forms of Eq. (\ref{fgdef}), with $a_f = 0.18, a_g = 0.07$. The exponents take very similar values using either fitting method. A. $P / P^*$ versus $N^{1/\nu} \tau$, with best-fit values $\nu=1.125$, $\beta = 0.50$, 95\% CI [0.47,0.53]. B. $S/ S^*$ versus $N^{1/\nu} \tau$, with best-fit values $\nu=1.125$, $\gamma = 0.13$, 95\% CI [0.11,0.15]. }
 \label{fig:4}
\end{figure}

These estimates yield predictions for the critical exponents $\beta$ and $\gamma$, given the value of $\nu$. In practice we investigate the dependence of $P$ and $S$ on the critical parameter, $\tau$, directly as we vary $\nu$, as in Figure 4. These quantities also scale as expected in the light of Eq. (\ref{PSfinite3}), and provide a consistency check that there is a single set of exponents that describes all fields at the transition.

Figure 5 shows the allowed region for these parameters as the intersection of the uncertainty ranges in the fits of Figs. 3-4, plus the dimensional limits arising from the hyperscaling relation of Eq. (\ref{hyperscaling}). We see that the allowed region corresponds to an effective  dimensionality $d \simeq 1$. The best-fit parameters are obtained at
\beq
\begin{split}
\nu = 1.125 \pm 0.10, &\> \beta = 0.50 \pm 0.03, \> \gamma = 0.13 \pm 0.02 \\
& \rightarrow d=1.00 \pm 0.08 ,
\end{split}
\label{critexpempirical1}
\eeq
where errors are computed at 95\% confidence. These estimates suffer from some uncertainties, as most fields show only limited and noisy scaling with $\tau$ in the critical region. Nevertheless, the fact that the critical dynamics of all nine fields may be fit reasonably well with a single set of critical exponents provides suggestive evidence for universality. Moreover, the fact that the field of carbon nanotubes --- a subfield of the large research area of nanotechnology --- shows the same critical dynamics near its transition point as its larger embedding field points to self-similarity: fields and subfields of research obey the same basic scaling relationships.

The estimate of a low dimensionality for the transition, $d \simeq 1$, suggests that near the critical point the networks are structured as strings of dense cliques, each clique weakly linked to the next like blobs on a line. Within each clique one finds high local degree, $k \gg 1$, with loose connections from a few nodes to the next neighboring clique. Large-scale connectivity --- of the sort that allows the entire network to percolate into a single connected component --- thus behaves essentially as a linear (one-dimensional) chain of collaboration. 

\begin{figure}
   \centering
   \includegraphics[width=3.2in]{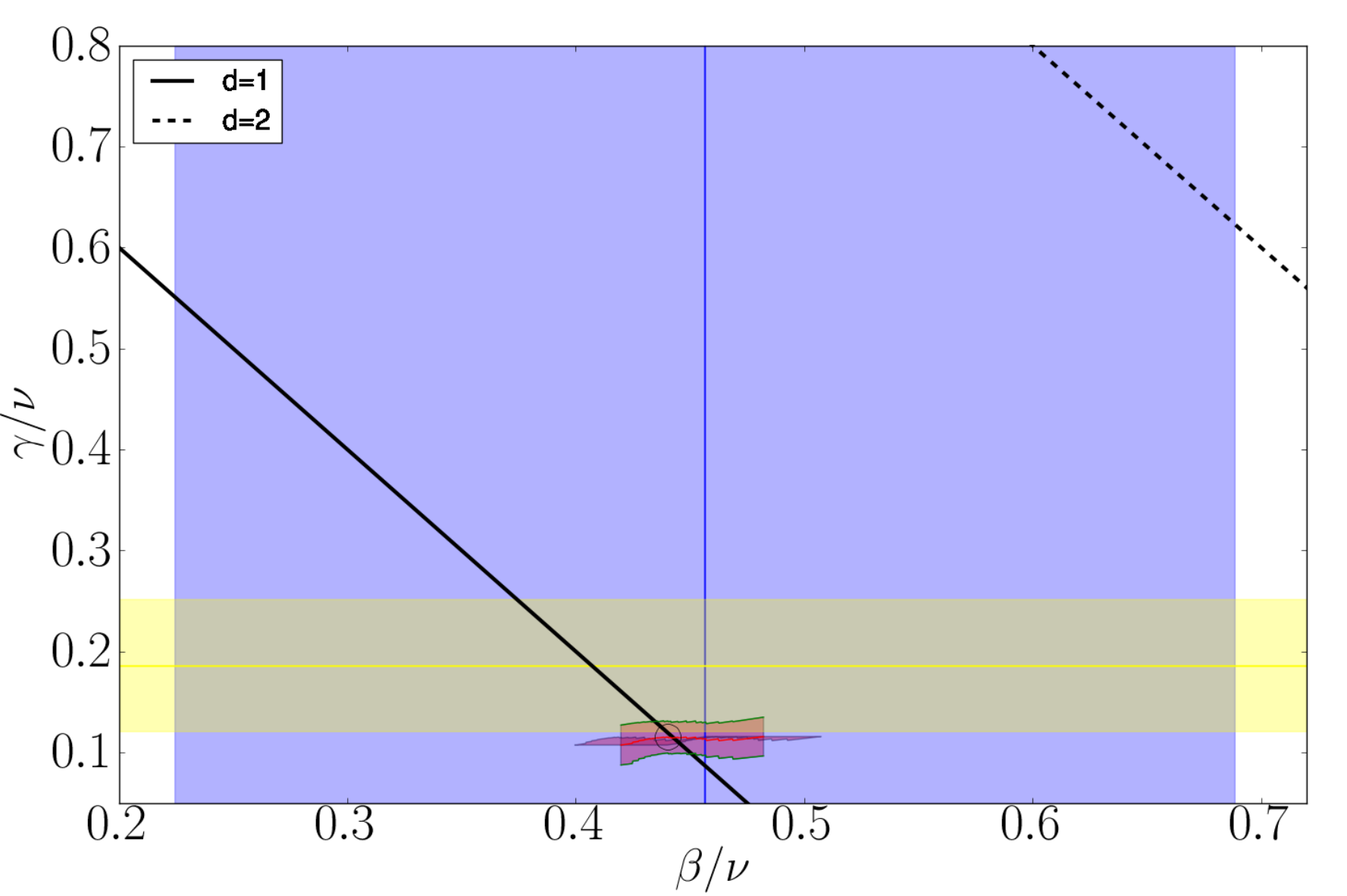} 
   \caption{\small The allowed parameter space of exponents from the finite-size scaling at criticality of $P$ and $S$ (blue and yellow regions, respectively) and direct estimate of their dependence on $\tau$ (red region).  Solid and dashed lines show dimensionality constraints on the exponent ratios resulting from the hyperscaling relation of Eq. (\ref{hyperscaling}). The allowed range for the ratios at 95\% confidence corresponds to the intersection of the yellow, blue and red regions, above the solid black line.  The best fits are obtained in the vicinity of $\nu \simeq 1.125$, with $\beta/\nu=0.44$ and $\gamma / \nu = 0.12$, $d=1.00$. These results suggest a common description of all fields as a percolation phenomenon dominated by chains of collaboration at criticality, corresponding to  an effective dimension $d \simeq 1$. The circle indicates the location of the best-fit parameters overall.}
   \label{fig:6}
\end{figure}

\subsection{Non-equilibrium phase transitions and the dynamics of discovery}

Our results suggest a low effective dimension, $d \simeq 1$, around the time that each field undergoes its topological transition. This finding is consistent with previous studies, such as collaboration networks in biotechnology \cite{biotech}, and indicates that around the time of field formation, distant authors are connected primarily by linear chains (trees) of collaboration strung between local clusters or cliques of co-authors. Because critical phenomena in $d = 1$ are simpler to analyze than higher-dimensional systems, these findings suggest that a single model might account for the observed critical dynamics of fields of any size. 

In our previous work \cite{BKKCW08,BCKC06,sustainability}, we found that for each of the scientific fields under study, the addition of new authors may be modeled as an epidemic contact process with a latency or ``incubation" period: after being ``exposed" to a given field, new authors require a nontrivial period of training before they become ``infected" and publish articles in the field themselves. This suggests that the network microdynamics might involve long-range interactions (such as temporal L\'{e}vy flights), which typically increase $\beta$ and reduce $\nu$ from their ordinary, mean-field values ($\beta_{\rm MF} = 1, \nu_{\rm MF} = 1/2$) \cite{Henkel,HinrichsenReview}. 

Over the past few years considerable progress has been made in identifying universality classes of non-equilibrium, stochastic spreading processes, including epidemic models with long-range interactions \cite{Henkel,HinrichsenReview}. Such models depart from the Directed Percolation (DP) universality class; instead, the critical exponents become functions of the parameters characterizing the interaction function. A new class of contact processes, which includes long-range interactions, was recently introduced in \cite{Ginelli,Argolo}, and is particularly relevant for our analysis.
In this model one considers a set of sites that can be either susceptible or infected. A spreading process occurs between an infected site and a nearby susceptible one at distance $l$, with probability $p(l) \sim l^{-a}$. These models thus describe a {\it density-dependent} form of interaction at a distance, which is different from the simplest form of long-range interactions. A recent numerical investigation found $\beta / \nu = 0.46$ and $\gamma / \nu = 0.10$ (and hence $d = 1.02$) for $a = 1.8$ in this family of models \cite{Argolo}, remarkably close to our best-fit parameters in Eq. (\ref{critexpempirical1}). 

In whatever form the effective interactions are ultimately expressed spatially, the 1D character of the field-formation transition becomes inevitable when we ask for the network structure that at once connects the graph globally (which defines the transition) and that is minimal in terms of the number of edges for a given number of nodes. Such a structure, which describes the large-scale structure of the field and should become ever more apparent as $N\rightarrow \infty$, is necessarily a {\it tree graph} thus resulting, in this limit, in $d\rightarrow 1$. 


\section{Discussion}

Scientific fields are self-organizing collections of people, their knowledge and interactions, and the physical products of their research. These collections evolve over time, seemingly in quite different ways. Scientific fields vary widely, for example, in the number of active researchers at any given time, the average number of co-authors per article, the rates at which new authors or articles join a field per year, and so on. When plotting ordinary time-series data for such quantities across scientific fields, diversity is the norm. 

These differences mask important underlying similarities in how scientific fields grow and develop. Such similarities become evident when we examine how various features of the fields' evolution scale not with ordinary time, $t$, but with number of authors, $N (t)$. 

We have focused on a particular type of scaling behavior in this paper. Building on earlier findings \cite{BKK09}, we have investigated a topological transition in co-authorship networks that tends to occur early in the development of new scientific fields. As new fields begin to cohere, researchers share common subject matter and research techniques, reinforced by topical conferences and journals. These features facilitate a certain kind of social dynamics: collaboration networks grow more densely connected over time, until the average connectivity per author crosses a critical threshold. Near that critical point, scientific fields undergo a topological transition, akin to percolation. Following the transition, nearly every author working in the field may be connected to every other author by a finite series of co-authorship links.

Just as with phase transitions in physical systems, some features of these patterns vary by field. The densification exponent, $\alpha_i$, critical connectivity, $k_c$, and network size at criticality, $N_c$, all differ among the nine scientific fields we have investigated (see Table 2). These quantities presumably depend sensitively upon microdynamics specific to each field, much as boiling points depend upon interatomic forces and hence differ widely across physical substances. 

Crucially, however, other features of the transition appear to be shared among each of the scientific fields under study. In particular, all of the fields we have examined --- which range from theoretical physics to biomedical research and huge, interdisciplinary research areas like sustainability science and nanotechnology, spanning several orders of magnitude in numbers of authors, articles, and rates of growth over time --- appear to share the same critical dynamics near their topological transitions. 

Under the usual scaling hypothesis, long familiar in statistical mechanics for accounting for finite-size effects in critical phenomena \cite{Fisher1971,ZinnJustin,Binder,Henkel}, we find suggestive evidence that all nine of the scientific fields under study are finite-sized realizations of a single, idealized, infinite-volume transition. All nine scientific fields, in other words, appear to be members of the same universality class, governed by a single set of critical exponents.

The scaling formalism assumes self-similarity: near the critical point, any portion of the system, if magnified to the size of the entire system, should behave indisinguishably from the entire system itself \cite{Fisher1971,ZinnJustin,Binder,Henkel}. Our data enable us to test the hypothesis of scale-invariance directly, by comparing the critical behavior of the fields of nanotechnology and its subspecialty, carbon nanotubes. Both the embedding field (nanotechnology) and its smaller subunit (carbon nanotubes) belong to the same universality class.



The existence of a general theory and detailed model that describes the formation of scientific fields across disciplines, time, and population size would provide a new comprehensive, quantitative, and predictive framework with which to understand the social and conceptual dynamics that drive the self-organized creation of scientific communities. Such a framework would be of significant interest to scientists and would hold great promise for guiding science policy.




\section{Materials}

On the search strategies used to construct the datasets for the first five fields listed in Table 1, see the appendix in \cite{BKKCW08}. To construct the database of all authors and articles in nanotechnology, we implemented the search strategy described in \cite{MogoutovNano} within the Web of Science database. The database of authors and articles on sustainability science was generated for a recent study of the evolution and structure of this field, see \cite{sustainability}.




\section{Acknowledgements}

This work was partially supported by the U.S. Department of Energy (DoE) Office of Scientific and Technical Information (OSTI); by the U.S. Department of Energy under grant Contract Number DE-SC00012567; and by the U.S. National Science Foundation (NSF) via grant SBE-0965259. We also gratefully acknowledge assistance from Jasleen Kaur, Shreeharsh Kelkar, and Andrei Mogoutov for help in constructing and parsing the database of publications on nanotechnology.

\end{document}